\documentstyle[prl,aps,multicol]{revtex}
\input psfig
\tighten
\begin{document}

\draft
\title{Tunneling ``zero-bias'' anomaly in the quasi-ballistic regime}
\author{A.M. Rudin, I.L. Aleiner, and L.I. Glazman}
\address{Theoretical Physics Institute, University of Minnesota, Minneapolis MN
55455}
\maketitle

\begin{abstract}
For the first time, we study the tunneling density of states (DOS) of the
interacting electron gas beyond the diffusive limit. A strong correction to the
DOS persists even at electron energies exceeding the inverse transport
relaxation time, which could not be expected from the well-known
Altshuler-Aronov-Lee (AAL) theory. This correction originates from the interference
between the electron waves scattered by an impurity and by the Friedel
oscillation this impurity creates. Account for such processes also revises the AAL
formula for the DOS in the diffusive limit.                                        
\end{abstract}
\pacs{PACS numbers: 73.40.Gk, 71.10.Pm}

\begin{multicols}{2}

\narrowtext
A number of experiments\cite{Imry} on tunneling in disordered
conductors consistently reveal a suppression of the differential conductance at
low biases, commonly referred to as zero-bias anomaly. This anomaly has found a
coherent explanation in the well-known Altshuler-Aronov-Lee (AAL)
theory\cite{Altshuler,Review} of one-particle density of states. AAL have shown that the
electron-electron interaction in the presence of disorder results
in a negative correction to the density of states (DOS), which is singular at Fermi
energy. In the case of a two-dimensional conductor, the AAL result reads:
\begin{equation} 
{\delta \nu^{AAL}(\epsilon) \over \nu} = A \ln (|\epsilon |\tau/\hbar),
\label{AAL}
\end{equation}
where the parameter $A>0$ depends on the details of the inter-electron
interaction, and is inversely proportional to the transport relaxation time
$\tau$. Correction (\ref{AAL}) diverges if the electron energy approaches the
Fermi level, $\epsilon\to 0$. The AAL theory assumes the diffusive motion of
electrons, which constraints the electron energy to the interval $\epsilon
<\hbar/\tau$. Clearly, in the case of a strong disorder, $\epsilon_F\tau\sim \hbar$,
this condition is not restrictive. However in cleaner samples (e.g.,
heterostructures with tunable density of two-dimensional electron
gas\cite{Eisenstein}) the energy domain
$\epsilon>\hbar/\tau$ becomes accessible, while the region of applicability of
Eq.~(\ref{AAL}) shrinks. 

One may infer from Eq.~(\ref{AAL}) that the correction to the DOS vanishes at
energies larger than $\hbar / \tau$, i.e., in the quasi-ballistic regime. The
main purpose of this Letter is to show that, to the contrary, interaction
does lead to a  significant correction to the DOS even for electron energies
exceeding the inverse transport relaxation time. This correction arises from the
interference of scattering on an impurity and on the Friedel oscillation it
creates. Account for such processes also revises the original AAL formula for
the DOS in the diffusive limit, adding a non-singular however large 
contribution to Eq.~(\ref{AAL}). 

Electron density of states  for energies larger than the inverse transport
relaxation  time is associated with the electron dynamics on the time
scale shorter than $\tau$. During such a small time an electron does not experience a
large number of scattering events, i.e., the scattering on disorder potential can
be treated in the lowest order of the perturbation theory, in contrast to
the diffusive limit. This approximation accounts only for the trajectories of electrons
which were scattered only once. We will show that the logarithmically divergent
correction to the density of states appears already in this approximation. 

We start with the most instructive case of the finite-range interaction potential,
and will calculate the correction to the one-particle DOS in the quasi-ballistic
limit due to a single short-range scatterer. Consider an impurity at the origin; its
potential $U_{\rm imp}({\bf r})$ induces a modulation of electron density around the
impurity. In the Born approximation, one can find the oscillating correction to the
electron density, $\delta n (r)$, which is known as the Friedel oscillation:
\begin{equation} \label{n(r)}
\delta n (r) = - {\nu g \over 2 \pi} {\sin (2 k_F r) \over r^2}. 
\end{equation}
Here $r$ is the distance from the impurity, $\nu = m/\pi\hbar^2$ is the
free-electron density of states, $m$ is the electron mass,
$k_F$ is the  Fermi wave vector, and  $g = \int U_{\rm imp}({\bf r}) d{\bf r}$.
In the presence of interaction $V({\bf r}-{\bf r'})$ between electrons, this density
oscillation produces an additional scattering potential, which
can be presented as a sum of  Hartree and exchange (Fock)  terms:
\begin{mathletters}\begin{eqnarray} \label{Hartree}
V_H ({\bf r}) &=& \int V({\bf r - r'}) \delta n({r'})d{\bf r'} \\
\label{exc}
V_{ex} ({\bf r}_1, {\bf r}_2) &=& V({\bf r}_1 - {\bf r}_2) 
\frac{\delta\rho ({\bf r}_1,{\bf r}_2)}{2},
\end{eqnarray}
where $\delta\rho({\bf r}_1, {\bf r}_2)\approx\delta n[(r_1+r_2)/2]$ is the
perturbation of the density matrix by the impurity. The exchange interaction
occurs only with the electrons of the same spin, which is reflected by the
factor $1/2$ in Eq.~(\ref{exc}). The Hartree-Fock energy (\ref{Hartree}), (\ref{exc})
oscillates as a function of distance from the impurity in the same manner as 
$\delta n(r)$ does. 
\label{HF}
\end{mathletters}

The local density of states is related to the retarded Green function,
$\nu (\epsilon, {\bf r}) = -(2/\pi)\, {\rm Im}G^R_{\epsilon} ({\bf r}, {\bf r})$. Let us
find now the correction $\delta G^R_{\epsilon} ({\bf r},  {\bf r})$ due to a coherent
process, which includes  a scattering on  the impurity potential itself, and a
scattering on the potential (\ref{HF}) formed by the Friedel oscillation:
\begin{eqnarray}\label{deltaG}
\delta &G^R_{\epsilon}& ({\bf r}, {\bf r}) = 2g \left\{
G^R_{\epsilon}({\bf r}) \int G^R_{\epsilon}({\bf r}_1)  V_H({\bf r}_1)
G^R_{\epsilon}({\bf r}_1 - {\bf r})  d{\bf r}_1 \right. \nonumber \\ 
& - & \left. G^R_{\epsilon}({\bf r}) \int G^R_{\epsilon}({\bf r}_1) 
V_{ex} ({\bf r}_1, {\bf r}_2) G^R_{\epsilon}({\bf r}_2 - {\bf r}) d{\bf r}_1 d{\bf
r}_2 \right\}. 
\end{eqnarray}
The Green function $G^R_{\epsilon}({\bf r})$ for a free electron 
at large distances, $k_Fr\gg 1$, and small energies, $\epsilon\ll\epsilon_F$,
is
\begin{equation} \label{G0} 
G^R_{\epsilon}({\bf r}) = {m e^{i\pi/4} \over
\hbar^2\sqrt{2\pi k_F r}}  e^{i(k_F + \epsilon /\hbar v_F) r}
\end{equation}
in two dimensions; $\epsilon$ is measured from the chemical potential.

Below we will be interested in the density of states averaged over the spatial scales
much larger than the Fermi wave length $\lambda_F\equiv 2\pi/k_F$. Therefore, we should
retain only those corrections $\delta\nu(\epsilon,{\bf r})$, which are smooth
functions of ${\bf r}$. Let us show now, using the Hartree contribution as an
example, that Eq.~(\ref{deltaG}) indeed yields such a correction. This contribution
corresponds to the following process. Electron starts motion in the point ${\bf r}$,
then experiences two scatterings,  (1) on the impurity potential in the origin, and
(2) on the  potential formed by the Friedel oscillation  in point ${\bf r}_1$,
and finally returns to point ${\bf r}$, see Fig.~\ref{fig1}. Motion along this
closed contour is represented in Eq.~(\ref{deltaG}) by the product
$G^R_{\epsilon}({\bf r}) G^R_{\epsilon}({\bf r}_1) G^R_{\epsilon}({\bf r}_1-{\bf
r})\propto \exp [i\phi({\bf r}, {\bf r}_1)]$, where  
\begin{equation}
\phi ({\bf r}, {\bf r}_1)=(r+r_1+|{\bf r}_1-{\bf r}|)(k_F +\epsilon/\hbar v_F)
\label{phase}
\end{equation}
is the geometric phase acquired by the electron. There is another strongly
oscillating factor in the integrand of Eq.~(\ref{deltaG}) -- the scattering potential 
$V_H({\bf r}_1)\propto\sin(2k_Fr_1)$. Obviously, the result of integration is
determined by the domain in space where the total phase of the integrand, 
$\phi ({\bf r}, {\bf r}_1)-2k_Fr_1$, is a slow function of ${\bf r}_1$. The
corresponding  electron trajectories are those  close  to the straight line, see
trajectory $A$ on Fig.~1. At $r_1>r$, Eq.~(\ref{phase}) yields the total phase of the
integrand $2(\epsilon/\hbar v_F)r_1$. Remarkably, this phase does not depend on
${\bf r}$. As a result, the Hartree correction to the Green function,
Eq.~(\ref{deltaG}), becomes  a {\it non-oscillating} function of $r$. Similar
arguments can be applied to the evaluation of the exchange correction to the Green
function. The resulting expression for the interaction correction to the local DOS is 
\begin{equation}\label{deltaNu1}
\delta \nu(\epsilon, {\bf r}) \approx  -{[V(0) - 2V(2k_F)] \nu^4 g^2 \over 8
k_F^2 r^2},
\end{equation}
if the distance from the impurity lies within the interval
$\max\{\lambda_F, d\}\lesssim r\lesssim\hbar v_F/\epsilon$, drops rapidly
($\propto 1/r^3$) at $r\gtrsim\hbar v_F/\epsilon$, and saturates at
$r\lesssim\max\{\lambda_F, d\}$. Here $d$ is the characteristic
spatial scale of the interaction potential, and $V(0)$ and
$V(2k_F)$ are the Fourier components of the interaction potential 
 appearing from the exchange and the Hartree terms respectively.

In order to find the averaged density of states one should sum up
contributions of the  type given by Eq.~(\ref{deltaNu1}) from all the 
impurities and then average over point ${\bf r}$ where the correction is
measured. Introducing the concentration of impurities $n_i$  and using
$\hbar/\tau=2\pi\nu n_i g^2$, we arrive to the following expression for the
interaction correction to the averaged DOS in quasi-ballistic ($\epsilon \gg
\hbar/\tau$) limit:
\begin{equation}\label{deltaNu}
{\delta \nu (\epsilon) \over \nu} \equiv 
-{\left< \delta \nu (\epsilon, {\bf r}) \right>
\over \nu} = { [V(0) -2V(2k_F)] \nu \hbar \over 4 \pi  \epsilon_F \tau}
\ln \left|\frac{\epsilon}{\Delta}\right|
\end{equation}
with $\Delta=\min \{\epsilon_F, \hbar v_F/d\}$. 

{\narrowtext
\begin{figure}[h]
\vspace{-0.5cm}
\hspace*{0.7cm}\psfig{figure=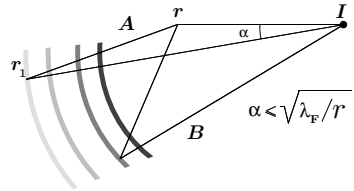,height=2.3in}
\vspace{-0.5cm}
\caption{Two typical trajectories ($A$, $B$) of an electron scattered by an
impurity ($I$) and by the corresponding Friedel oscillation (concentric arcs). The
correction $\delta\nu ({\bf r})$ is dominated by the trajectories of the type
 $A$, for which  the electron is almost scattered back at $I$ and ${\bf r}_1$.} 
\label{fig1}
\end{figure}
}

In principle, the correction can have any sign depending on the relation between
$V(0)$ and $V(2k_F)$. However, in any realistic system  the
inter-electron interaction is sufficiently smooth, $d \gtrsim \lambda_F$, and
$V(0)\gg V(2k_F)$. Therefore, in the following we will concentrate on the 
exchange contribution, which always dominates in the correction to the DOS.

The derivation of Eq.~(\ref{deltaNu}) is valid for energies $\epsilon$ exceeding
$\hbar/\tau$, which is the high-energy cutoff in the AAL theory, see Eq.~(\ref{AAL}).
Quasi-ballistic formula (\ref{deltaNu}) at the boundary of the region
of its applicability, $\epsilon\sim\hbar/\tau$, does not match the AAL result
(\ref{AAL}). The reason of this mismatch is the choice of the high-energy
cutoff. Physically, AAL cutoff  means taking into account only electron states
within a narrow energy strip, $-\hbar/\tau\lesssim\varepsilon <0$, below the Fermi
level, in the formation of  the Friedel oscillation. This cut-off was dictated by the
range of applicability of the diffusion approximation for the electron dynamics  AAL
used\cite{Altshuler,Review}. On the other hand, our analysis leading to
Eq.~(\ref{deltaNu}) demonstrates that electron states within a
much wider strip, $-\Delta\lesssim\varepsilon <0$, are important for the
correction. As it turns out, this wider strip is important for the calculation
of the DOS at $\epsilon\lesssim\hbar/\tau$ as well.  To show this and to remedy
the mismatch, below we calculate the DOS without using the diffusion
approximation.

We are interested in the spatially averaged density of states, which makes it
possible to use the standard diagrammatic techniques. The correction to
the averaged one-particle density of states has the form:
\begin{equation} \label{nu1}
\delta \nu (\epsilon, T) = 
-{2\over \pi}\,  {\rm Im} \int {d {\bf p} \over (2\pi)^2} \, \delta G(i
\epsilon_n \rightarrow \epsilon + i0, {\bf p}),
\end{equation}  
where $\epsilon_n=\pi(2n+1)T$ is Matsubara frequency, $T$ is the temperature.
(For brevity we  omit the Planck constant  in all the intermediate formulas.) We will
calculate  the correction to the electron propagator, 
$\delta G(i\epsilon_n, {\bf p})$, in the first order in the
screened electron-electron interaction $V_{sc}(i\Omega_l, {\bf Q})$. In the
metallic regime ($\epsilon_F\tau\gg 1$) the exchange contribution to the
propagator is:
\begin{eqnarray} \label{dG}
& \delta & G(i\epsilon_n, {\bf p}) = -[ G(i\epsilon_n, {\bf p})]^2 T \sum_{
\Omega_l} \int  {d{\bf Q} \over (2\pi)^2}  \, \theta(\epsilon_n(\Omega_l -
\epsilon_n)) \nonumber \\ 
& & \times [\Gamma( i \Omega_l, {\bf Q})]^2 
V_{sc}(i\Omega_l, {\bf Q})  G(i\epsilon_n - \Omega_l, {\bf p-Q}).
\end{eqnarray}
Here $G(i\epsilon_n, {\bf p}) = [i\epsilon_n - \xi_p + (i/2\tau)\, {\rm
sign}\epsilon_n]^{-1}$ is the electron Green's function in the dirty
conductor, and $\Gamma$ is the vertex function calculated in the ladder
approximation.  As long  as we are developing theory applicable for
any relation between electron energy and $\hbar/\tau$, we cannot use the usual
diffusion form for vertex function. The formula valid for an arbitrary momentum ${\bf
Q}$ and energy $\Omega_l$ transfer is:
\begin{equation} \label{Gamma}
\Gamma(i\Omega_l, {\bf Q}) = \left( 1 - { 1/\tau \over \sqrt{(|\Omega_l |+ 1/\tau)^2 +
(v_F Q)^2}}\right)^{-1}.
\end{equation}
Note, that in the limit $\Omega_l, v_F Q\ll\tau^{-1}$, Eq. (\ref{Gamma}) reduces
to the standard diffusion expression. 

Integration over ${\bf p}$ in Eq. (\ref{nu1}) gives
\begin{eqnarray} \label{nu2}
{\delta \nu (\epsilon, T)  \over \nu} & = & -{\rm Re} \, \lim _{i\epsilon_n
\rightarrow 
\epsilon + i0}T \sum_{\Omega_l} \int  {d{\bf Q} \over (2\pi)^2} 
\theta(\epsilon_n(\Omega_l - \epsilon_n)) 
 \nonumber \\
& \times & {  [\Gamma( i \Omega_l, {\bf Q})]^2 
V_{sc}(i\Omega_l, {\bf Q})  \,  (|\Omega_l | + 1/\tau) \over \left[(|\Omega_l
| + 1/\tau) ^2 +  (v_F Q)^2 \right]^{3/2}}.
\end{eqnarray}

The case of a finite-range electron-electron  interaction is especially
simple, because we can replace $V_{sc}(i\Omega_l, {\bf Q})$ in Eq. (\ref{nu2}) by
the Fourier component of the unscreened interaction potential $V(Q)$. In
this case the correction derived from Eq.(\ref{nu2}) coincides with the exchange
term in formula (\ref{deltaNu}). It means that the exchange correction to the DOS
\begin{equation}\label{short}
{\delta \nu (\epsilon) \over \nu}  = {V(0) \nu \hbar \over 4 \pi 
\epsilon_F \tau}
\ln \left|\frac{\epsilon}{\Delta}\right|
\end{equation}
is universal, i.e., is valid for the energies  both larger  and smaller than
$\hbar/\tau$.

For the long-range Coulomb interaction, $V(Q)=2\pi e^2/Q$, the screening  should
be taken into account, $V_{sc}(i\Omega_l, {\bf Q}) = V (Q) /[1+ V (Q) \Pi(i\Omega_l,
Q)]$. Here the polarization operator
\begin{equation} \label{Pi}
\Pi(i\Omega_l, {\bf Q}) = \nu  \left(1 - { \Gamma (i\Omega_l, Q) 
|\Omega_l| \over  \sqrt{(|\Omega_l |+ 1/\tau)^2 + (v_F Q)^2}}\right)
\end{equation}
is  derived in the random phase approximation. 

Straightforward evaluation  of Eq. (\ref{nu2}) with the account for 
Eqs.~(\ref{Gamma})  and (\ref{Pi}) yields
\begin{equation} \label{nu3}
{\delta \nu (\epsilon, T)  \over \nu} = - {1 \over 8\pi  \epsilon_F \tau} 
 \int _{\bar{\epsilon}}^{\Delta}
{d\Omega \over  \Omega}  \ln \left({v_F^2 /a_B^2\over 
 \Omega \sqrt{\Omega^2 + (1/\tau)^2}} \right) , 
\end{equation}
where $a_B=1/me^2$ is the Bohr radius, $\bar{\epsilon} \equiv\max\{|\epsilon|, T\}$,
and the cut-off energy $\Delta$ is given now by $\Delta = \hbar v_F/a_B$. 

Equation (\ref{nu3}) gives the  correction to the
one-particle DOS due to the Coulomb interaction and is the main  quantitative
result of this Letter.   The integral in Eq.~(\ref{nu3}) cannot be expressed in terms
of elementary functions. However, the diffusive ($\epsilon \ll\hbar/\tau$) 
and quasiballistic ($\epsilon \gg \hbar/\tau$) asymptotic behaviors are easily found.

In the diffusive limit  exchange correction  to the  one-particle DOS  has the 
form:
\begin{eqnarray} \label{nu4}
{\delta \nu_{dif} (\bar{\epsilon})  \over \nu} &= &- {\hbar \over  16\pi
\epsilon _F\tau} \left\{  \ln 
\left({\bar{\epsilon} \over \hbar D^2 a_B^{-4} \tau } \right) \ln (\bar{\epsilon} 
\tau/\hbar)  \right. \nonumber \\
& + & \left.  2\left[ \ln(\tau \Delta /\hbar)\right]^2\right\},
\end{eqnarray}
where $D=v_F^2 \tau/2$ is the diffusion coefficient. The first term of the sum
in (\ref{nu4}) is the famous result of the Altshuler-Aronov-Lee theory
\cite{Altshuler}.  The second, new term is not singular. This part of
the correction represents the contribution of electrons deep in the Fermi sea, with
energies below the ``$\hbar/\tau$ strip''.

{\narrowtext
\begin{figure}[h]
\hspace*{-0.8cm}\psfig{figure=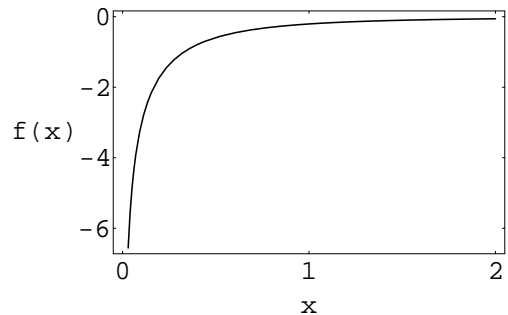,height=1.9in}
\vspace{0.4cm}
\caption{The crossover function $f(x)$, see Eq. (\protect\ref{nu6}).} 
\label{fig2}
\end{figure}
}

In the quasi-ballistic limit  exchange correction  to the  one-particle DOS is: 
\begin{equation} \label{nu5} 
{\delta \nu_{bal} (\bar{\epsilon})  \over \nu} = -
{\hbar \over  8\pi  \epsilon _F \tau} \,  \left[ \ln (\bar{\epsilon} / \Delta )
\right]^2.
\end{equation} 
The leading term in the energy dependence of the correction is $\propto
(\ln\bar{\epsilon})^2$ at any relation between $\bar{\epsilon}$ and $\tau$.
In the crossover region, the correction is described by the formula
\begin{equation} \label{nu6}
\frac{\delta \nu(\bar{\epsilon})}{\nu} = \frac{\delta \nu_{bal} (\bar{\epsilon})}{\nu} -
\frac{\hbar}{8 \pi  \epsilon _F \tau}f(\bar{\epsilon}\tau).
\end{equation} 
The dimensionless function $f(x)$ is plotted in Fig.~2. Its asimptotes,
$f(x\to\infty)=0$ and $f(x\to 0)= -(1/2)[\ln (x)]^2$, allow one to obtain from
(\ref{nu6}) the limits (\ref{nu5}) and (\ref{nu4}) respectively.

In the practically important case of a  gated semiconductor structure, the
Coulomb potential is suppressed at all distances exceeding the
separation $d$ between the two-dimensional electron gas and the gate.
If $d\gg a_B$, then the correction to the density of states
coincides with Eq. (\ref{nu5}) at energies ${\bar \epsilon} >\hbar v_F/\sqrt{a_B d}$, and
becomes logarithmical\cite{foot} at smaller energies,
\begin{equation} \label{nuZuzin}
{\delta \nu (\bar{\epsilon})  \over \nu} = {\hbar \over  8\pi \epsilon _F 
\tau} \,   \ln (d/a_B)  
\, \ln (\bar{\epsilon} /\Delta),
\end{equation}
with $\Delta = \hbar v_F/a_B$. This formula is applicable for an arbitrary
relation between $\epsilon$ and $\hbar/\tau$. 

The density of states Eq.~(\ref{nu1}) describes adequately the electron tunneling
without the lateral momentum conservation, such as tunneling through an inhomogeneous
barrier. However, the electron-electron interaction affects the tunneling through a
homogeneous barrier as well. We will consider below tunneling between two identical
quantum wells (QW), assuming (in accordance with the experiments\cite{Eisenstein}) the
lateral momentum conservation in the course of tunneling. The correction to the
conductance we find, has a logarithmic zero-bias anomaly. 

In the absence of disorder and of the electron-electron interaction, the
conservation of in-plane momentum implies that an electron can tunnel only if the levels
of spatial quantization in the wells line up precisely. This makes differential tunneling
conductance of a symmetrical double-QW system have a singular peak at zero bias.
Disorder alone smears the singularity, leading to the $I$-$V$
characteristic\cite{Zheng}
\begin{equation}\label{current1}
I_0(V) = G_0 {eV\, (\hbar/\tau_s)^2 \over (eV)^2 + (\hbar/\tau_s)^2},
\end{equation}
with the width given by the inverse quantum lifetime of electrons in the wells, $1/\tau_s$.
Here $V$ is the bias applied to the contact, and $G_0$ is the zero bias conductance.

Electron-electron interaction adds a singular at zero bias, negative logarithmic
correction to the current. Details of the calculations will be published elsewhere; the
result for the interaction correction to the tunneling current is: 
\begin{equation}\label{current}
{\delta I(V) \over I_0(V)} \approx { \hbar \over \pi \epsilon_F \tau} \, 
\ln(d/a_B)\, 
\ln(\overline{eV} / \Delta).
\end{equation}
Here $\overline{eV} \equiv \max \{eV, T\}$ is assumed to satisfy \cite{footnote2} the
conditions $\overline{eV} \ll
\hbar/\tau_s$, $v_F/\sqrt{a_B d}$, and $d$ is the separation between wells. 
 Qualitatively Eq. (\ref{current}) and Eq.~(\ref{nuZuzin}) have the
same feature: the upper  cut-off for the correction is $\Delta=\hbar v_F/a_B$, and by no
means $\hbar /\tau$. The $\propto\ln eV$ bias  dependence, instead of $\propto (\ln eV)^2$
one, appears in Eq.~(\ref{current}) because of a partial cancellation\cite{Shytov} of the
corrections coming from the intra- and inter-well   electron-electron  interaction. In the
absence of interaction, Eq.~(\ref{current1}) would lead to a peak in the differential
conductance $dI/dV$ at zero bias. Negative diverging correction (\ref{current}) splits this
peak in two.  The separation between the  maxima of these  two sub-peaks is:
\begin{equation}\label{eVs}
eV_{sp} = \frac{\hbar}{\tau_s}\sqrt{\frac{\ln (d/a_B)}{8\pi\epsilon_F \tau/\hbar}}.
\end{equation}
The sub-peaks should be resolved at sufficiently low temperatures, $T\lesssim eV_s$.
Estimate for $V_{sp}$ for the data of Turner {\it et al}, Ref. \onlinecite{Eisenstein},
gives $V_{sp}\approx 0.05$mV. It is important, that Eqs. (\ref{current}) and (\ref{eVs})
are valid at any relation between $eV, eV_{sp}$,  and the energy $\hbar /\tau$.

In summary,  we studied the tunneling density of states  of interacting
two-dimensional electron gas beyond the diffusive limit. A significant
interaction-induced suppression of the density of states persists at electron energies
even larger than the inverse transport relaxation time, which could not be expected from
the well-known Altshuler-Aronov-Lee theory\cite{Altshuler}. The AAL formula  for the
density of states at low energies is also revised, and an additional non-singular, 
however  large, contribution was found. 

Discussions with B.L. Altshuler and A.I. Larkin are acknowledged with gratitude. This work
was supported  by NSF Grant DMR-9423244.

\end{multicols}
\end{document}